\documentclass[5p,twocolumn,superscriptaddress]{revtex4-2}

\usepackage{lineno}
\usepackage{xcolor}
\usepackage{graphicx} 
\usepackage{subfigure}
\usepackage{dcolumn} 
\usepackage{bm}
\usepackage{float}
\usepackage{hyperref}
\usepackage{amsmath}

\begin{document}

\title{Effect of vector meson spin coherence on the observables for the chiral magnetic effect in heavy-ion collisions}
\author{Zhiyi Wang}
\affiliation{Key Laboratory of Nuclear Physics and Ion-beam Application (MOE), Institute of Modern Physics, Fudan University, Shanghai 200433, China}
\affiliation{Shanghai Research Center for Theoretical Nuclear Physics, NSFC and Fudan University, Shanghai 200438, China}
\author{Jinhui Chen}
\affiliation{Key Laboratory of Nuclear Physics and Ion-beam Application (MOE), Institute of Modern Physics, Fudan University, Shanghai 200433, China}
\affiliation{Shanghai Research Center for Theoretical Nuclear Physics, NSFC and Fudan University, Shanghai 200438, China}
\author{Diyu Shen}
\email{dyshen@fudan.edu.cn}
\affiliation{Key Laboratory of Nuclear Physics and Ion-beam Application (MOE), Institute of Modern Physics, Fudan University, Shanghai 200433, China}
\affiliation{Shanghai Research Center for Theoretical Nuclear Physics, NSFC and Fudan University, Shanghai 200438, China}
\author{Aihong Tang}
\affiliation{Brookhaven National Laboratory, Upton, New York 11973, USA}
\author{Gang Wang}
\affiliation{Department of Physics and Astronomy,
  University of California, Los Angeles, California 90095, USA}
\date{\today}

\begin{abstract}
The chiral magnetic effect (CME) in heavy-ion collisions reflects the local violation of ${\cal P}$ and ${\cal CP}$ symmetries in strong interactions and manifests as electric charge separation along the direction of the magnetic field created by the wounded nuclei. The experimental observables for the CME, such as the $\gamma_{112}$ correlator, the $R_{\Psi_2}(\Delta S)$ correlator, and the signed balance functions, however, are also subject to non-CME backgrounds, including those from resonance decays. A previous study showed that the CME observables are affected by the diagonal component of the spin density matrix, the $\rho_{00}$ for vector mesons. In this work, we study the contributions from the other elements of the spin density matrix using a toy model and a multiphase transport model. We find that the real part of the $\rho_{1-1}$ component, $\mathrm{Re}\,\rho_{1-1}$, affects the CME observables in a manner opposite to that of the $\rho_{00}$. All three aforementioned CME observables show a linear dependence on $\mathrm{Re}\,\rho_{1-1}$ in the model calculations, supporting our analytical derivations. The rest elements of the spin density matrix do not contribute to the CME observables. The off-diagonal terms in the spin density matrix indicate spin coherence and may be nonzero in heavy-ion collisions due to local spin polarization or spin-spin correlations. Thus, $\mathrm{Re}\,\rho_{1-1}$, along with $\rho_{00}$, could play a significant role in interpreting measurements in search of the CME.

\end{abstract}

\maketitle

\section{Introduction}
Experiments in heavy-ion collisions at Relativistic Heavy Ion Collider (RHIC) and the Large Hadron Collider (LHC) can create a deconfined nuclear matter known as quark-gluon plasma (QGP)~\cite{Shuryak:1980tp,BRAHMS:2004adc,PHOBOS:2004zne,STAR:2005gfr,PHENIX:2004vcz,Busza:2018rrf,Chen:2018tnh}, providing a unique test ground for quantum chromodynamics (QCD). Particularly, the chiral magnetic effect (CME) probes the topological vacuum transition in QCD~\cite{Kharzeev:2007jp,Fukushima:2008xe}, whereby ${\cal P}$ and ${\cal CP}$ symmetries may be locally violated in strong interactions. The CME predicts electric charge separation along the magnetic field generated in heavy-ion collisions~\cite{Voronyuk:2011jd,Deng:2012pc,STAR:2023jdd}.
Several experimental observables have been proposed to detect the CME-induced charge separation, such as the ${\gamma }_{112}$ correlator~\cite{22Voloshin:2004vk}, the $R_{\Psi_2}(\Delta S)$ correlator~\cite{23Magdy:2017yje}, and signed balance functions~\cite{20Tang:2019pbl}, with their core components found to be equivalent~\cite{24Choudhury:2021jwd}. The search for the CME using these observables has been extensively conducted at RHIC and the LHC over the past two decades~\cite{7STAR:2009wot,8STAR:2013ksd,9STAR:2014uiw,10STAR:2021mii,11STAR:2022ahj,12ALICE:2012nhw,13CMS:2016wfo,14CMS:2017lrw,15ALICE:2017sss,STAR:2023ioo,Chen:2024zwk,Kharzeev:2024zzm}. However, a firm conclusion on the existence of the CME in such experiments remains elusive, as data interpretation is impeded by the incomplete understanding of non-CME backgrounds, especially those related to the collective motion or elliptic flow ($v_2$) of the collision system~\cite{16Liao:2010nv,17Pratt:2010zn,18Wang:2016iov,19Feng:2018chm,21Wu:2022fwz,Xu:2023elq}. 

Besides the flow-related backgrounds, our prior study showed that the 00-component of the spin density matrix for vector mesons, the $\rho_{00}$, could also affect the CME observables because of the anisotropic decay pattern of two oppositely charged daughters~\cite{DShen:plb}. In such studies, the spin density matrix is defined along the direction perpendicular to the reaction plane (spanned by impact parameter and beam momenta), 
\begin{equation}
\rho^{V}=\left(\begin{array}{c c c}{{\rho_{11}}}&{{\rho_{10}}}&{{\rho_{1-1}}}\\ {{\rho_{01}}}&{{\rho_{00}}}&{{\rho_{0-1}}}\\ {{\rho_{-11}}}&{{\rho_{-10}}}&{{\rho_{-1-1}}}\end{array}\right),    
\label{eq:rhoV}
\end{equation}
where the indices 1, 0, and $-1$ label the vector meson's spin components along the spin-quantization axis. The deviation of $\rho_{00}$ from 1/3 with respect to the reaction plane is known as the global spin alignment effect ~\cite{25Liang:2004ph,26Liang:2004xn}, which has been measured in experiment for several vector mesons, such as $\phi$, ${K}^{*0}$~\cite{42STAR:2008lcm,43ALICE:2019aid,Nature}, and $J/\Psi$~\cite{ALICE:2022dyy}. The observed global spin alignment effects for those mesons are unexpectedly large compared to predictions based on  the global polarization of $\Lambda$ hyperons~\cite{37STAR:2017ckg,38STAR:2018gyt,39ALICE:2019onw,40STAR:2021beb}, suggesting rich physics mechanisms beyond spin-orbital coupling, e.g., the strong electromagnetic fields~\cite{STAR:2023jdd,28Yang:2017sdk,29Sheng:2019kmk}, local spin alignment~\cite{30Xia:2020tyd,31Gao:2021rom}, as well as novel phenomena such as fluctuation of strong vector meson fields~\cite{29Sheng:2019kmk,34Sheng:2020ghv,42Sheng_prl,Chen:2023hnb,43Sheng_prd,Chen:2024afy} and/or local axial charge currents~\cite{32Muller:2021hpe}. All those mechanisms will cause $\rho_{00}$ to deviate from 1/3, resembling an apparent charge-separation. Therefore, the background contribution from $\rho_{00}$ to the CME observables cannot be ignored.

The off-diagonal elements in the spin density matrix reflect the spin coherence between the states 1, 0, and $-1$, and could be finite due to various physics effects. For example, local spin alignment~\cite{30Xia:2020tyd} can result in a nonzero real part of the $\rho_{1-1}$ component, the $\mathrm{Re}\,\rho_{1-1}$, in central collisions. Spin-spin correlations~\cite{Lv:2024uev} could also lead to nonzero off-diagonal elements, depending on the difference in the correlation strength between different directions. Finite $\mathrm{Re}\,\rho_{1-1}$ values have been observed in lepton-induced reactions and hadron-hadron collisions, such as those for ${K}^{*0}$ in $e^+e^-$ collisions at the Large Electron–Positron Collider (LEP)~\cite{OPAL:1997vmw}.
In this article, we adopt the framework from Ref.~\cite{DShen:plb} and study the impact of all components of the spin density matrix on the CME observables through the decay of $\rho \rightarrow \pi^+\pi^-$. For each of the aforementioned CME observables, we first perform an analytical derivation and then use a toy model and a multiphase transport model (AMPT)~\cite{Lin:2004en} to confirm the findings. 

\section{The \texorpdfstring{$\gamma_{112}$}{Lg} correlator}

The CME-induced electric dipole breaks the up-down symmetry across the reaction plane, resulting in nonzero sine terms in the azimuthal angle distribution of final-state particles~\cite{22Voloshin:2004vk},
\begin{equation}
\frac{dN_\pm}{d \varphi} \propto  1+ 2a_1^\pm\sin \Delta \varphi + \sum_{n=1}^{\infty} 2v_n^\pm\cos (n\Delta \varphi) ,
\end{equation}
where $\Delta \varphi = \varphi - \Psi_{\rm RP}$ is the azimuthal angle of a particle relative to the reaction plane ($\Psi_{\rm RP}$). 
$a_1^\pm$ characterizes the strength of charge separation with opposite signs for oppositely charged particles. $v_n^\pm$ denotes the $n^{\rm th}$-harmonic flow coefficient of final-state particles, with $v_2$ conventionally denoting the elliptic flow.

Since ${\cal P}$ is only locally violated but globally conserved, the event average of $a_1^\pm$ is zero. The CME has to be detected through fluctuations of charged particles. The $\gamma_{112}$ correlator  serves this purpose~\cite{22Voloshin:2004vk}, 
\begin{equation}
\gamma_{112} \equiv \left \langle \cos(\varphi_\alpha + \varphi_\beta - 2\Psi_{\rm RP}) \right \rangle,
\end{equation}
where $\varphi_\alpha$ and $\varphi_\beta$ are the azimuthal angles of charged particles $\alpha$ and $\beta$, respectively. 
The bracket denotes averaging over all particle pairs and all events. The CME signal is contained in the difference between the opposite-sign (OS) and same-sign (SS) pairs,
\begin{equation}
\Delta \gamma_{112} \equiv \gamma_{112}^{\rm{OS}} - \gamma_{112}^{\rm{SS}}\approx 2|a_1^\pm|^2.
\end{equation}
However, $\Delta \gamma_{112}$ is contaminated with charge-dependent backgrounds, while charge-independent ones are canceled out. We express the contribution from the decay of $\rho \rightarrow \pi^{+} + \pi^{-}$ as described in Ref.~\cite{DShen:plb}, 
\begin{eqnarray}
\Delta \gamma_{112}^\rho = \frac{N_\rho}{N_+N_-}
& & \big[\rm{Cov}( \cos \Delta \varphi_+, \cos \Delta \varphi_-) \notag \\
& & - \rm{Cov}( \sin \Delta \varphi_+, \sin \Delta \varphi_- )\big], 
\label{Eq: gamma}
\end{eqnarray}
where Cov$(a,b)$ denotes the covariance of variables $a$ and $b$. $N_\rho$ is the yield of $\rho$ meson, and $N_+$ and $N_-$ are the numbers of $\pi^+$ and $\pi^-$, respectively.  

\begin{figure}[htbp]
\centering
\vspace*{-0.1in}
\includegraphics[width=0.5\textwidth]{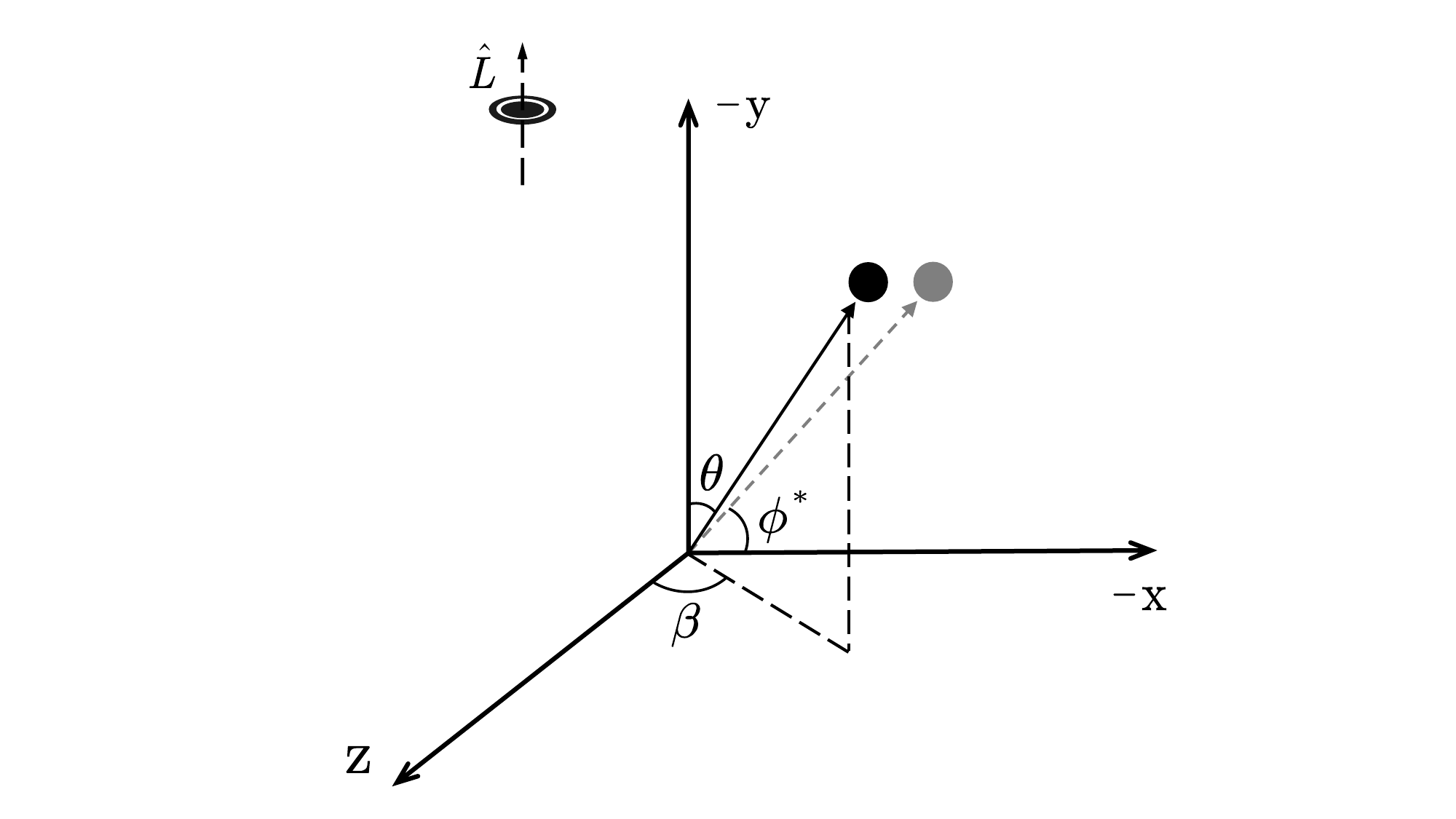}
\caption{\label{fig:Coordinate}
Sketch of pion emission (solid arrow) in the rest frame of the parent $\rho$ meson. The impact parameter and the beam direction are aligned along the $x$ and $z$ axes, respectively.
The dashed arrow represents the projection in the $x$-$y$ plane.}
\label{fig1}
\end{figure}

Figure~\ref{fig1} illustrates pion emission in the rest frame of the parent $\rho$ meson, where 
$\theta$ is the angle between the pion momentum ($\vec{p}$) and the system total angular momentum $\hat{L}$, $\phi^*$ is the azimuthal angle, and $\beta$ is the angle between the $z$ axis and the projection of $\vec{p}$ onto the $x$-$z$ plane.
Incorporating all elements of the spin density matrix, we write the distribution of the $\pi^\pm$ emission angle as outlined in Ref.~\cite{30Xia:2020tyd},
\begin{eqnarray}
\frac{d^{2}N}{d(\cos\theta)d\beta} &=& \frac{3}{8\pi}\big[\: (1-\rho_{00})+(3\rho_{00}-1)\cos^{2}\theta \notag\\ 
& &-\sqrt{2}\mathrm{Re}(\rho_{10}-\rho_{0-1})\sin(2\theta)\cos\beta \notag\\
& &+\sqrt{2}\mathrm{Im}(\rho_{10}-\rho_{0-1})\sin(2\theta)\sin\beta \notag\\
& &-2\mathrm{Re}\,\rho_{1-1}\sin^{2}\theta\cos(2\beta) \notag\\
& &+2\mathrm{Im}\,\rho_{1-1}\sin^{2}\theta\sin(2\beta) \:\big].
\label{Eq:density-theta/beta}
\end{eqnarray}
The projection of Eq.~(\ref{Eq:density-theta/beta}) to the transverse plane ($x$-$y$) becomes
\begin{eqnarray} 
\frac{dN}{d\phi^{*}} &=&\frac{1}{2\pi}\big[ 1-\frac{1}{2}(3\rho_{00}-1)\cos 2\phi^{*}  \notag\\
& &+\sqrt{2}\mathrm{Im}(\rho_{10}-\rho_{0-1})\sin 2\phi^{*} + \mathrm{Re}\,\rho_{1-1}\cos 2\phi^{*} \big]. \;\;\;\;
\label{Eq:density-phi}
\end{eqnarray}
Then, the covariance terms in Eq.~(\ref{Eq: gamma})  can be calculated in the rest frame of the $\rho$ meson as
\begin{eqnarray}
\rm{Cov}(\cos \phi_+^*, \cos \phi_-^* )
  &=& -\left \langle \cos^2 \phi_+^* \right \rangle + \left \langle \cos \phi_+^* \right \rangle^2  \notag \\
 &=& -\frac{1}{2}+\frac{3\rho_{00}-1}{8}-\frac{\mathrm{Re}\,\rho_{1-1}}{4},  \label{Eq:Covcc} \\
\rm{Cov}( \sin \phi_+^*, \sin\phi_-^* ) 
  &=& -\left \langle \sin^2 \phi_+^* \right \rangle + \left \langle \sin \phi_+^* \right \rangle^2  \notag \\
 &=& -\frac{1}{2}-\frac{3\rho_{00}-1}{8}+\frac{\mathrm{Re}\,\rho_{1-1}}{4}.  \label{Eq:Covss}
\end{eqnarray} 
Here we take $\phi_-^* = \phi_+^* +  \pi$. Therefore,  the decay contribution $\Delta \gamma_{112}^\rho$ in Eq.~(\ref{Eq: gamma}) becomes
\begin{equation}
\Delta \gamma_{112}^{\rho*} = \frac{N_\rho}{N_+N_-} \left[\: \frac{3}{4} (\rho_{00}-\frac{1}{3} ) - \frac{1}{2}\mathrm{Re}\,\rho_{1-1} \:\right],
\label{Eq:rest_result}
\end{equation}
in the $\rho$ rest frame.
Apparently, both $\rho_{00}$ and ${\rm Re}\,\rho_{1-1}$ could contribute to the $\Delta\gamma_{112}$ correlator, whereas all other elements of the spin density matrix can be safely ignored.

Back in the laboratory frame, Eqs.~(\ref{Eq:Covcc}) and (\ref{Eq:Covss}) need to be scaled by factors of $f_c$ and $f_s$, respectively, due to the Lorentz boost of the $\rho$ meson~\cite{DShen:plb}. 
In general, $f_c$ and $f_s$ are different because of the anisotropic collective motion ($v_2^\rho$) of $\rho$ mesons. Since the elliptic flow effect has been discussed previously~\cite{DShen:plb}, we assume $v_2^\rho$ is zero for simplicity. Hence, the background contribution of decay pions to $\Delta \gamma_{112}$ in the laboratory frame can be expressed as
\begin{equation}
\Delta \gamma_{112}^{\rho} = f_0 \frac{ N_\rho }{N_+N_-} \left[\: \frac{3}{4}  (\rho_{00}-\frac{1}{3} ) - \frac{1}{2}\mathrm{Re}\,\rho_{1-1} \:\right],
\label{Eq:gamma_lab}
\end{equation}
where the coefficient $f_0$ absorbs the Lorentz boost effects and should depend on the 
spectrum of the $\rho$ meson~\cite{DShen:plb}. 

\begin{figure}[htbp]
	\centering
{\includegraphics[scale=0.33]{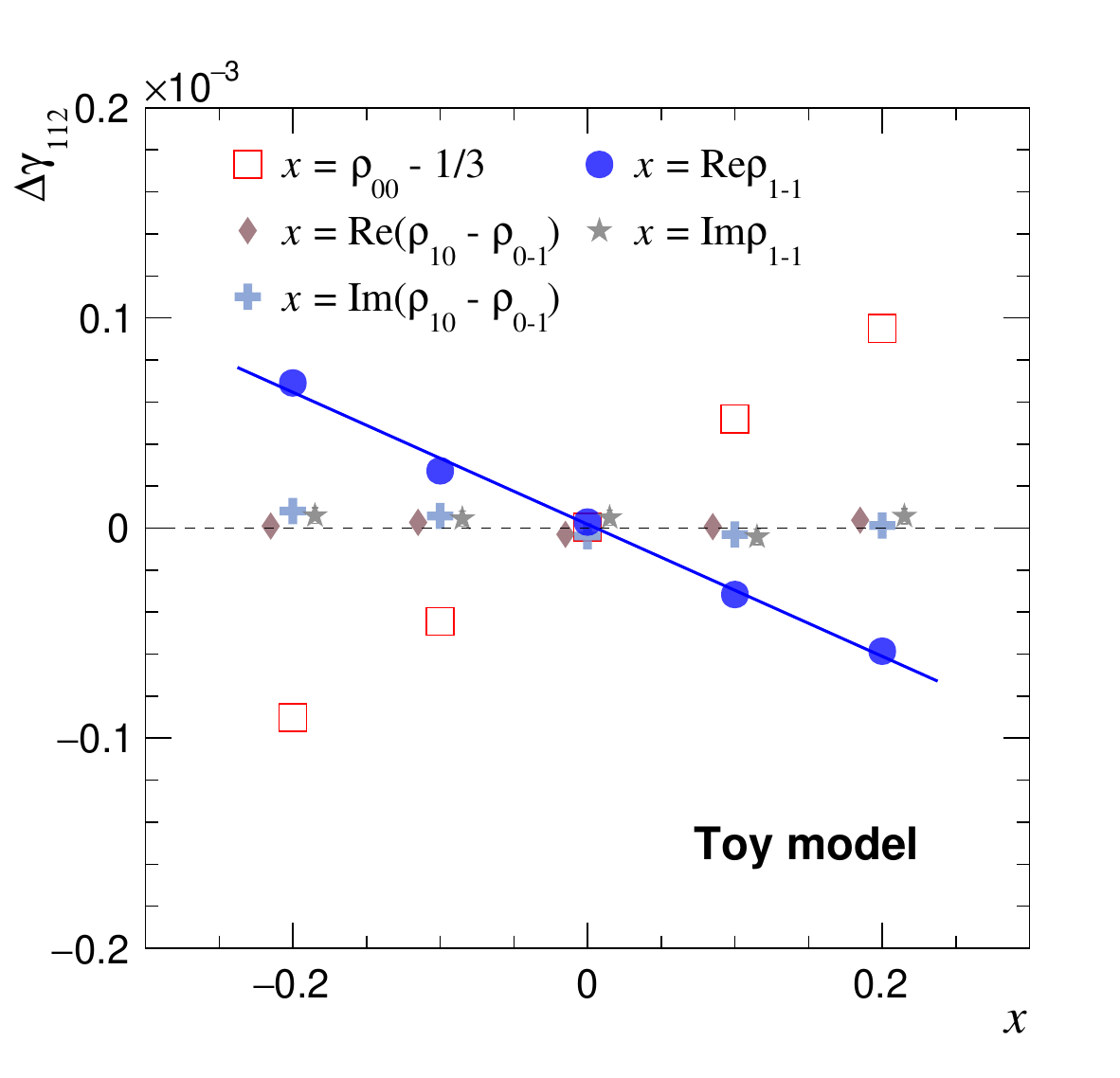}}
\caption{Toy model simulations of the $\pi$-$\pi$ $\Delta \gamma_{112}$ correlation vs various elements of the $\rho$-meson spin density matrix: $\rho_{00}-1/3$, ${\rm Re}(\rho_{10}-\rho_{0-1})$, ${\rm Im}(\rho_{10}-\rho_{0-1})$, ${\rm Re}\,\rho_{1-1}$, and ${\rm Im}\,\rho_{1-1}$.
$v_2^\rho$ is set to zero.  The solid line represents the linear fit to the case $x = {\rm Re}\,\rho_{1-1}$.}
\label{fig:Gamma_toy}
\end{figure}

We first test the analytical derivation using toy model simulations without the CME. We follow the same implementations as in the previous study~\cite{DShen:plb}, updating the momentum distribution of $\rho$-meson decay products according to Eq.~(\ref{Eq:density-theta/beta}). We set $v_2^\rho$ to be zero.
Figure~\ref{fig:Gamma_toy} shows the simulation results of the $\pi$-$\pi$ $\Delta \gamma_{112}$ correlation as a function of $\rho_{00}-1/3$, ${\rm Re}(\rho_{10}-\rho_{0-1})$, ${\rm Im}(\rho_{10}-\rho_{0-1})$, ${\rm Re}\,\rho_{1-1}$, and ${\rm Im}\,\rho_{1-1}$, respectively. $\Delta \gamma_{112}$ is linearly correlated with $\rho_{00}$ but anti-correlated with ${\rm Re}\,\rho_{1-1}$, supporting Eq.~(\ref{Eq:gamma_lab}), where $\rho_{00}$ and ${\rm Re}\,\rho_{1-1}$ have comparable but opposite impacts on $\Delta \gamma_{112}$. Other components have no contributions to $\Delta\gamma_{112}$.  

\begin{figure}[htbp]
	\centering
{\includegraphics[scale=0.33]{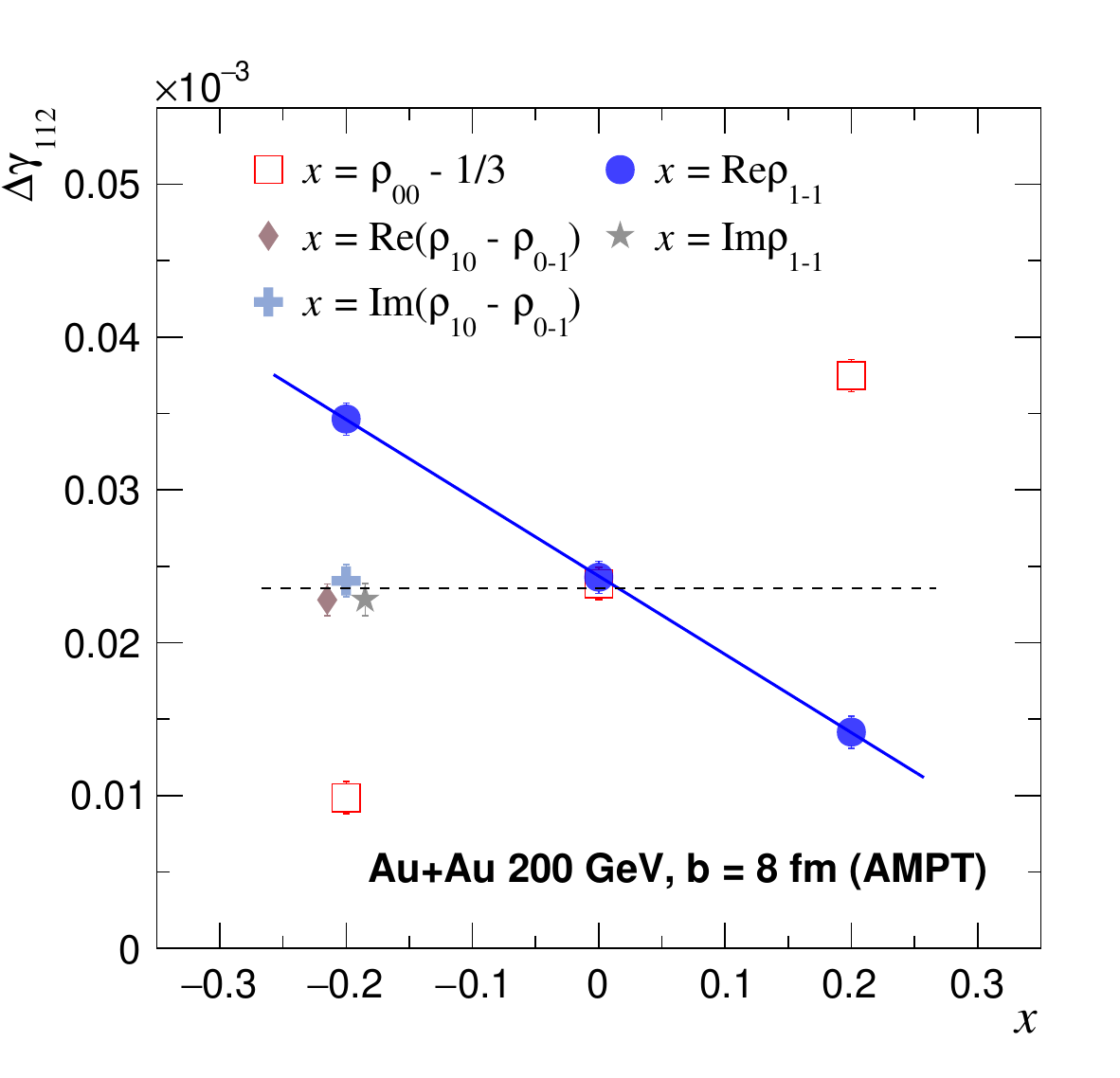}} 
\caption{AMPT calculations of the $\pi$-$\pi$ $\Delta \gamma_{112}$ correlation vs various elements of the $\rho$-meson spin density matrix: $\rho_{00}-1/3$, ${\rm Re}(\rho_{10}-\rho_{0-1})$, ${\rm Im}(\rho_{10}-\rho_{0-1})$, ${\rm Re}\,\rho_{1-1}$, and ${\rm Im}\,\rho_{1-1}$ in  Au+Au collisions at $\sqrt{s_{NN}}=200$ GeV with an impact parameter of 8 fm. The solid line denotes the linear fit for the case $x = {\rm Re}\,\rho_{1-1}$, while the dashed line represents a constant to guide the eye for cases involving non-contributing elements.}
	\label{fig:Gamma_ampt}
\end{figure}
We also study this effect using AMPT simulations of Au+Au collisions at $\sqrt{s_{NN}}=200$ GeV with an impact parameter of 8 fm. The details of this model can be found in Ref.~\cite{Lin:2004en}. The string-melting version of the AMPT model is used without the CME, and the selected decay channel is $\rho \rightarrow \pi^+ + \pi^-$. Pions are analyzed without any kinematic cut to increase statistics.
The spin density matrix is implemented by redistributing the momenta of decay products according to Eq.~(\ref{Eq:density-theta/beta}), similar to the procedures in Refs.~\cite{50Lan:2017nye,51Shen:2021pds}. 
The $\rho_{00}-1/3$ and Re$\,\rho_{1-1}$ values are each set to be $-0.2$, 0, and 0.2 to test the linear dependence. The other elements are set to $-0.2$ solely to demonstrate their contribution or lack thereof. For each case, we have generated 2 million events.
Figure~\ref{fig:Gamma_ampt} shows the linear correlation (anti-correlation) between $\Delta \gamma_{112}$ and $\rho_{00}$ (${\rm Re}\,\rho_{1-1}$), akin to the toy model simulations.
For other spin density matrix components,
the $\Delta \gamma_{112}$ values at $x = -0.2$  are consistent with that at $x=0$ within uncertainties, as demonstrated by a constant dashed line.
The nonzero $\Delta \gamma_{112}$ value at $x= 0$ comes from the positive $v_2^\rho$ and local charge conservation~\cite{17Pratt:2010zn,18Wang:2016iov}. 
Quantitatively, the slopes $d \Delta \gamma_{112}/d\rho_{00}$ and $d \Delta \gamma_{112}/d\mathrm{Re}\,\rho_{1-1}$ depend on the $\rho$-meson spectrum.

\section{The \texorpdfstring{$R_{\Psi_2}(\Delta S)$}{Lg} correlator}
Another CME observable, the $R_{\Psi_2}(\Delta S)$ correlator, focuses on the event-by-event fluctuations of $a_1$ and is defined as a double ratio of four distributions,
\begin{equation}
R_{\Psi_2}(\Delta S) \equiv \frac{N(\Delta S_{{\rm real}})}{N(\Delta S_{{\rm shuffled}})} / \frac{N(\Delta S^\perp_{{\rm real}})}{N(\Delta S^\perp_{{\rm shuffled}})} , \label{Eq:R2} 
\end{equation}
where
\begin{eqnarray}
\Delta S &=& \left \langle \sin (\phi^+ - \Psi_2)\right \rangle - \left \langle \sin (\phi^- - \Psi_2)\right \rangle,  
\label{Eq:R3} \\
\Delta S^\perp &=& \left \langle \cos (\phi^+ - \Psi_2)\right \rangle - \left \langle \cos (\phi^- -\Psi_2)\right \rangle,  
\label{Eq:R4}
\end{eqnarray}
where $\phi^+$ and $\phi^-$ are the azimuthal angles of positively and negatively charged particles, respectively. $\Psi_2$ represents the $2^{\rm nd}$-order event plane, which serves as a proxy of the reaction plane and is estimated from the final-state hadron emission. The bracket denotes averaging over all particles in an event. The subscript ``real" refers to using the actual charge information, while ``shuffled" indicates reshuffling charges within the same event. Ideally, the CME-driven charge separation should cause a concave shape in $R_{\Psi_2}(\Delta S)$. By construction, the resulting concave or convex shape relies on the relative widths of the four individual $N(\Delta S)$ distributions. Therefore, as in our previous work~\cite{DShen:plb}, we introduce a straightforward variable to quantify the signal or background strength based on the four variances,
\begin{eqnarray}
\Delta \sigma_{R}^2 
&=& \sigma^2(\Delta S_{\rm real}) - \sigma^2(\Delta S_{\rm shuffled})  \notag \\
& &- \sigma^2(\Delta S_{\rm real}^{\perp}) + \sigma^2(\Delta S_{\rm shuffled}^{\perp}). 
\label{Eq:deltaR}
\end{eqnarray}
The CME signal corresponds to a positive $\Delta \sigma_{R}^2$.

Again, we set $v_2^{\rho}$ to zero so that the Lorentz boost factor $f_0$ only depends on the $\rho$ spectrum. Then, each term of Eq.~(\ref{Eq:deltaR}) in the laboratory frame can be expressed as
\begin{flalign}
&\sigma^2(\Delta S_{\rm real}) = f_0\left[ \sigma_s^2 - \frac{2N_\rho}{N_+N_-} {\rm Cov}( \sin\Delta \phi_+, \sin\Delta \phi_- ) \right], \label{Eq:VarYReal} \\
&\sigma^2(\Delta S_{\rm real}^{\perp}) = f_0\left[ \sigma_c^2 - \frac{2N_\rho}{N_+N_-} {\rm Cov}(\cos\Delta \phi_+, \cos\Delta \phi_-) \right], \label{Eq:VarXReal} \\
&\sigma^2(\Delta S_{\rm shuffled}) =  f_0 \sigma_s^2,  \label{Eq:VarYShuf} \\
&\sigma^2(\Delta S_{\rm shuffled}^{\perp}) = f_0 \sigma_c^2,& \label{Eq:VarXShuf}
\end{flalign}
where
\begin{eqnarray}
\sigma_s^2 = \frac{\sigma^2 (\sin \phi_+^* )}{N_+}  + \frac{\sigma^2 (\sin \phi_-^*)}{N_-}, \\
\sigma_c^2 = \frac{\sigma^2(\cos \phi_+^*) }{N_+}  + \frac{\sigma^2 (\cos \phi_-^*)  }{N_-}. 
\end{eqnarray}
The shared terms will be canceled out, leaving only the covariance terms in $\Delta \sigma_{R}^2$.
According to Eqs.~(\ref{Eq:Covcc}) and (\ref{Eq:Covss}), the contribution of $\rho$-decay pions to $\Delta \sigma_{R}^2$ can be calculated as
\begin{equation}
\Delta \sigma_{R}^2 = f_0 \frac{ N_\rho}{N_+N_-} \left[ \: \frac{3}{2} ( \rho_{00}-\frac{1}{3} ) - \mathrm{Re}\,\rho_{1-1} \:\right].
\label{Eq:RCor_delta}    
\end{equation}
Eq.~(\ref{Eq:RCor_delta}) has a function form similar to that of Eq.~(\ref{Eq:gamma_lab}), with $\Delta \sigma_{R}^2$ also exhibiting a linear dependence on $\rho_{00}$ and Re$\,\rho_{1-1}$.

Alternatively, the CME signal can also be reflected in the difference between the inverse Gaussian widths,
\begin{eqnarray}
\frac{S_{\rm concavity}}{\sigma_{R}^2}  &=& \frac{1}{ \sigma^2(\Delta S_{\rm real})} - \frac{1}{\sigma^2(\Delta S_{\rm shuffled})}  \notag \\
& &- \frac{1}{\sigma^2(\Delta S_{\rm real}^{\perp})} + \frac{1}{\sigma^2(\Delta S_{\rm shuffled}^{\perp})}. 
\label{Eq:sigma_R2} 
\end{eqnarray}
$S_{\rm concavity}$ is 1 $(-1)$ when the $R_{\Psi_2}(\Delta S)$ distribution is convex (concave). After applying Eqs.~(\ref{Eq:VarYReal}, \ref{Eq:VarXReal}, \ref{Eq:VarYShuf}, \ref{Eq:VarXShuf}), we have
\begin{equation}
S_{\rm concavity} = {\rm Sign}\left[ \: \mathrm{Re}\,\rho_{1-1}- \frac{3}{2}(\rho_{00}-\frac{1}{3}) \: \right].
\label{Eq:sign}
\end{equation}
At $\rho_{00}=1/3$, a negative (positive) Re$\,\rho_{1-1}$ value corresponds to $S_{\rm concavity} = -1$ (1) and a concave (convex) $R_{\Psi_2}(\Delta S)$ distribution.

\begin{figure}[htbp]
	\centering
{\includegraphics[scale=0.33]{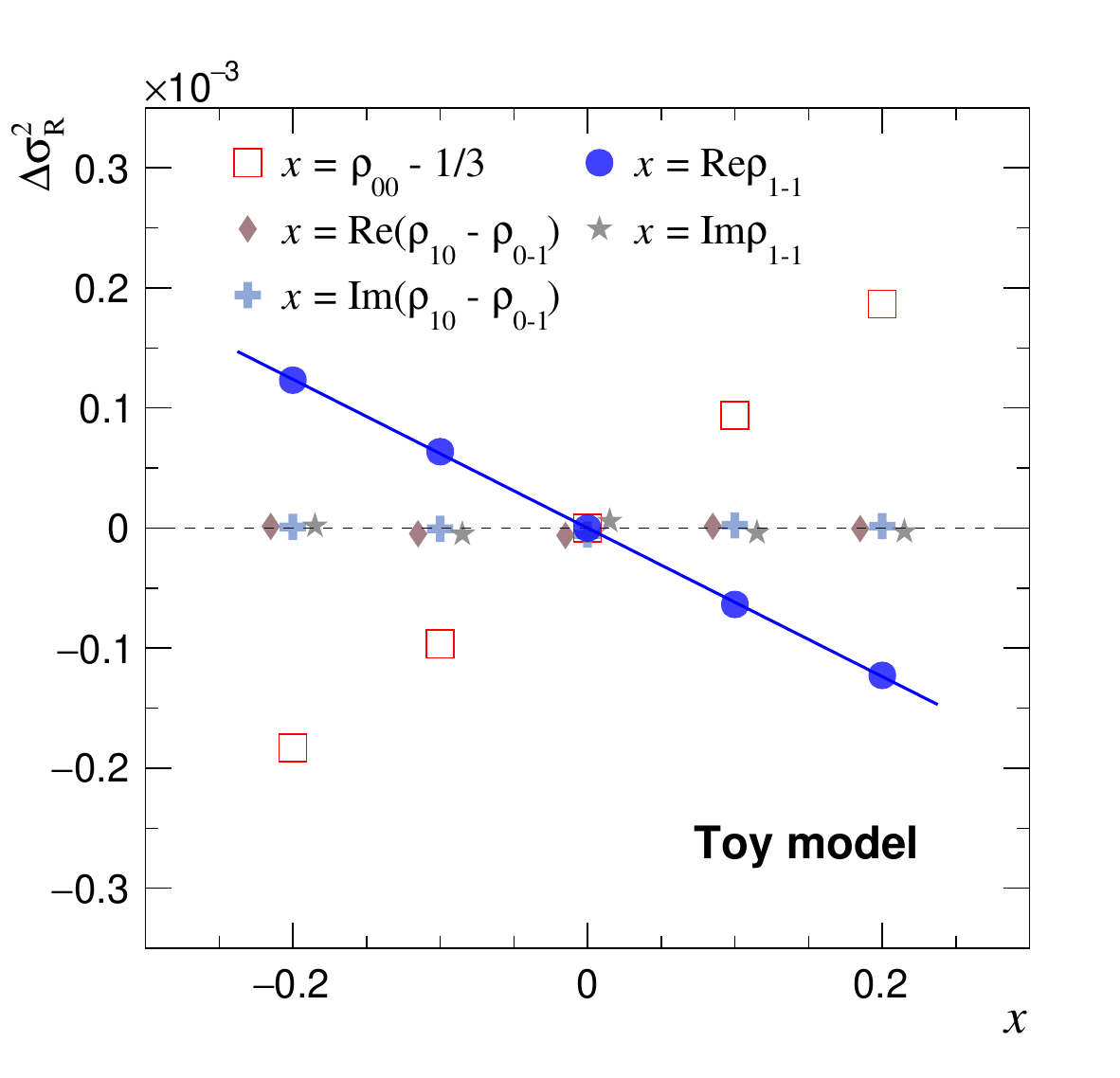}} 	
\caption{Toy model simulations of $\Delta \sigma_{R}^2$ vs various elements of the $\rho$-meson spin density matrix: $\rho_{00}-1/3$, ${\rm Re}(\rho_{10}-\rho_{0-1})$, ${\rm Im}(\rho_{10}-\rho_{0-1})$, ${\rm Re}\,\rho_{1-1}$, and ${\rm Im}\,\rho_{1-1}$.
$v_2^\rho$ is set to zero.  The solid line represents the linear fit to the case $x = {\rm Re}\,\rho_{1-1}$.}
\label{fig:deltaR_toy}
\end{figure}

Figure~\ref{fig:deltaR_toy} shows the toy-model simulations results of $\Delta \sigma_{R}^2$ as a function of each element of the $\rho$-meson spin density matrix,
with the same settings as those for the $\Delta\gamma_{112}$ calculations. As expected, $\Delta \sigma_{R}^2$ exhibits a linear dependence on $\rho_{00}$ and Re$\rho_{1-1}$, similar to the behavior of $\Delta\gamma_{112}$ in Fig.~\ref{fig:Gamma_toy}.  This indicates that, in addition to global spin alignment, the off-diagonal elements resulting from spin coherence also influence the $R_{\Psi_2}(\Delta S)$ correlator. The other elements do not contribute to the $\Delta \sigma_{R}^2$.

\begin{figure*}[htbp]
	\centering
{\includegraphics[width=0.4\linewidth]{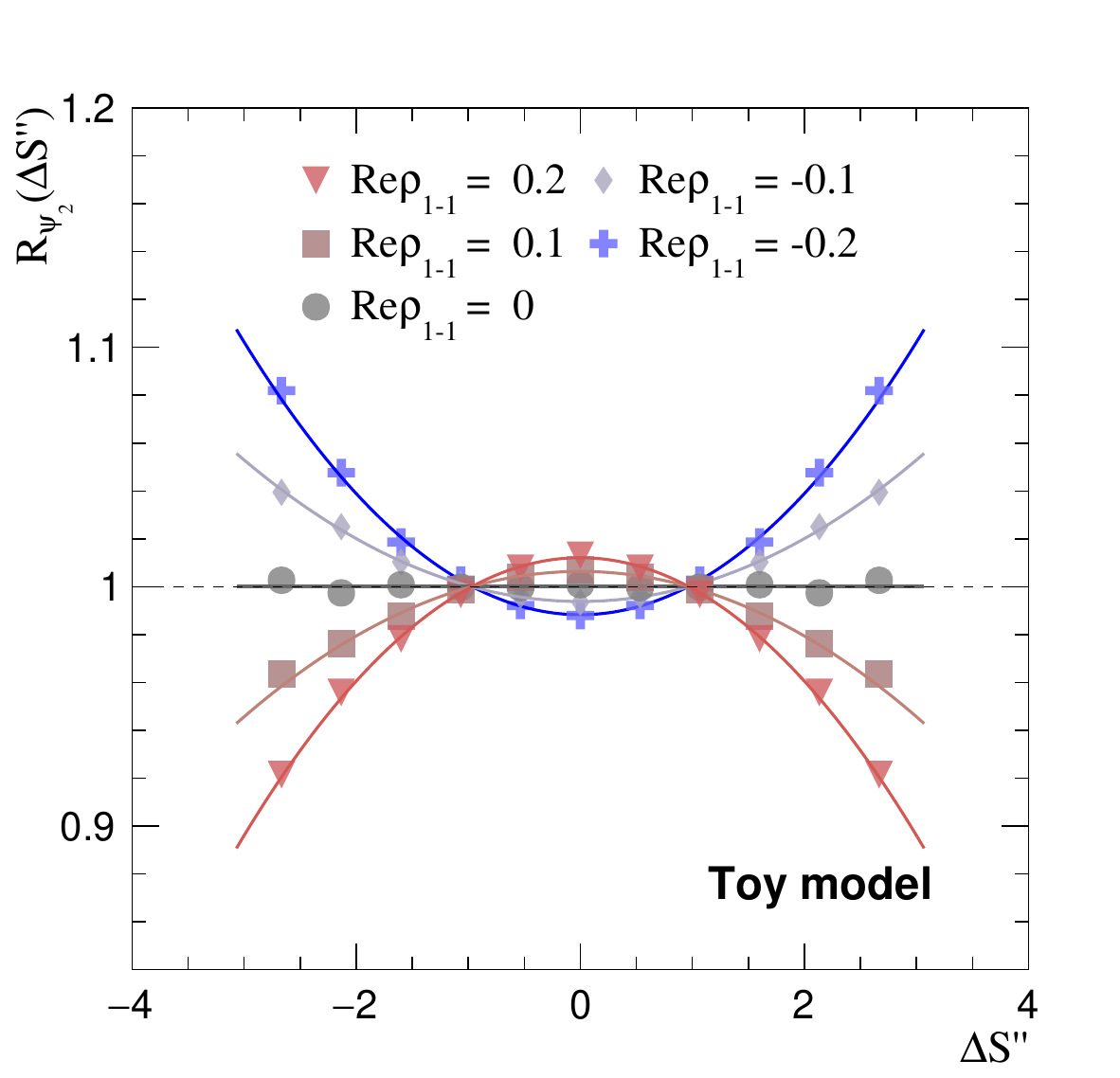}} 
{\includegraphics[width=0.4\linewidth]{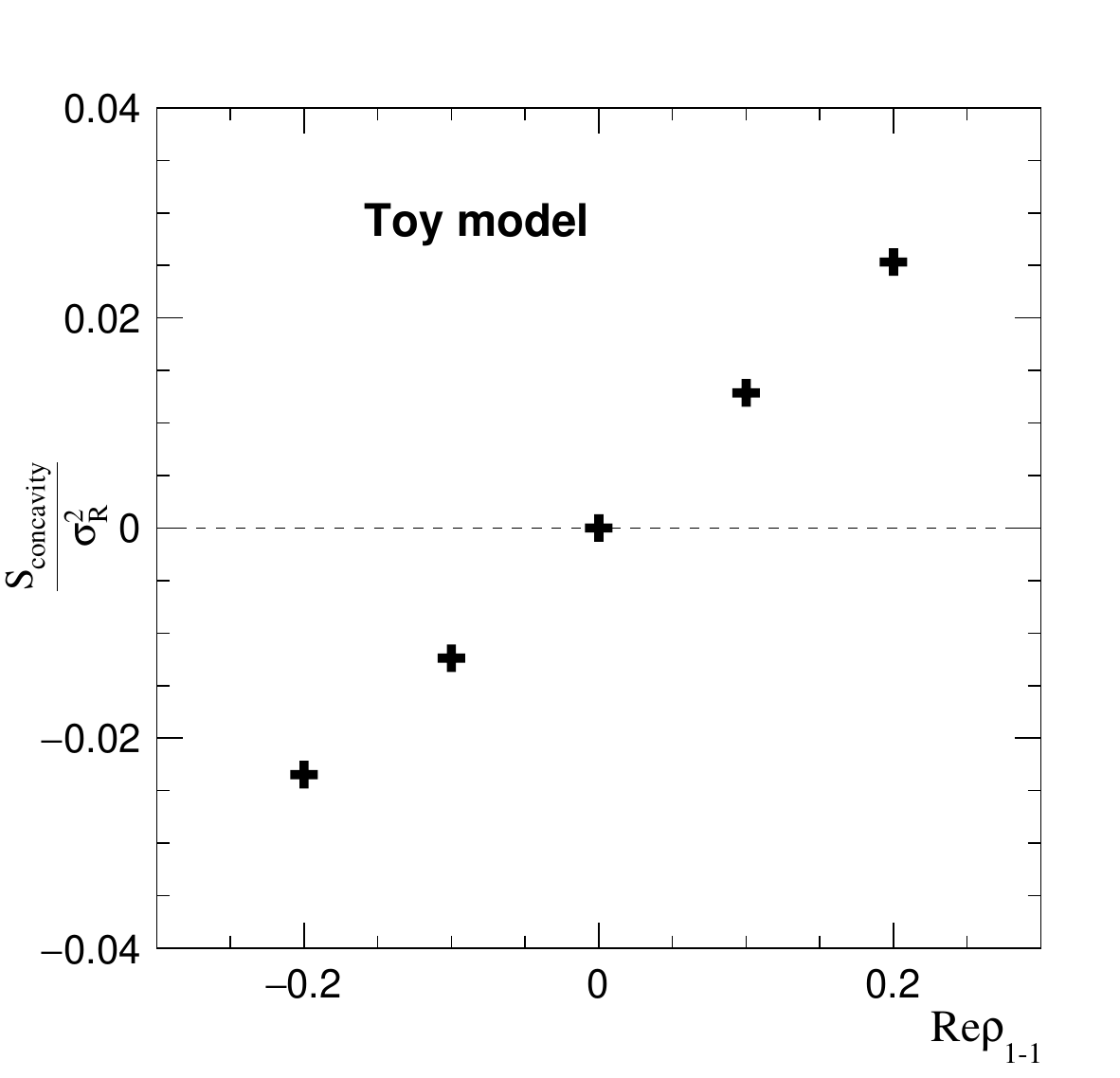}} 
\caption{(Left) toy model simulations of the $R_{\Psi_2}(\Delta S'')$ distribution with several Re$\,\rho_{1-1}$ inputs. $v_2^\rho$ and $\rho_{00}$ are set to 0 and $1/3$, respectively. The distributions are symmetrized around $\Delta S''=$0.  (Right) The $S_{\rm concavity}/\sigma_{R}^2$ values extracted using Gaussian fits to the $R_{\Psi_2}(\Delta S'')$ distributions at different Re$\,\rho_{1-1}$. }
\label{fig:RCor_toy}
\end{figure*}
To check the shape of the $R_{\Psi_2}(\Delta S)$ distribution, we adopt the same procedure as described in Ref.~\cite{23Magdy:2017yje} to correct for the effect of multiplicity fluctuations, by defining $\Delta S'' = \Delta S/\sigma_{\rm sh}$. Here, $\sigma_{\rm sh}$ represents the width of $N(\Delta S_{\rm shuffuled})$. 
Figure~\ref{fig:RCor_toy} (left) shows the $R_{\Psi_2}(\Delta S'')$ distributions with several Re$\,\rho_{1-1}$ values from the toy model simulations, where $v_2^\rho$ and $\rho_{00}$ are set to 0 and $1/3$, respectively. Re$\,\rho_{1-1}$ varies from $-0.2$ to 0.2. When Re$\,\rho_{1-1}$ is positive (negative), the $R_{\Psi_2}(\Delta S'')$ distribution exhibits a convex (concave) shape. Moreover, a larger magnitude of Re$\,\rho_{1-1}$ corresponds to a narrower $R_{\Psi_2}(\Delta S'')$ distribution. 
Figure~\ref{fig:RCor_toy} (right) shows the $S_{\rm concavity}/\sigma_{R}^2$ values extracted using Eq.~(\ref{Eq:sigma_R2}) for different Re$\,\rho_{1-1}$ inputs, with a clear rising trend. $S_{\rm concavity}$ appears to bear the same sign as Re$\,\rho_{1-1}$, which supports the analytical derivation.   

\begin{figure*}[htbp]
	\centering
{\includegraphics[width=0.4\linewidth]{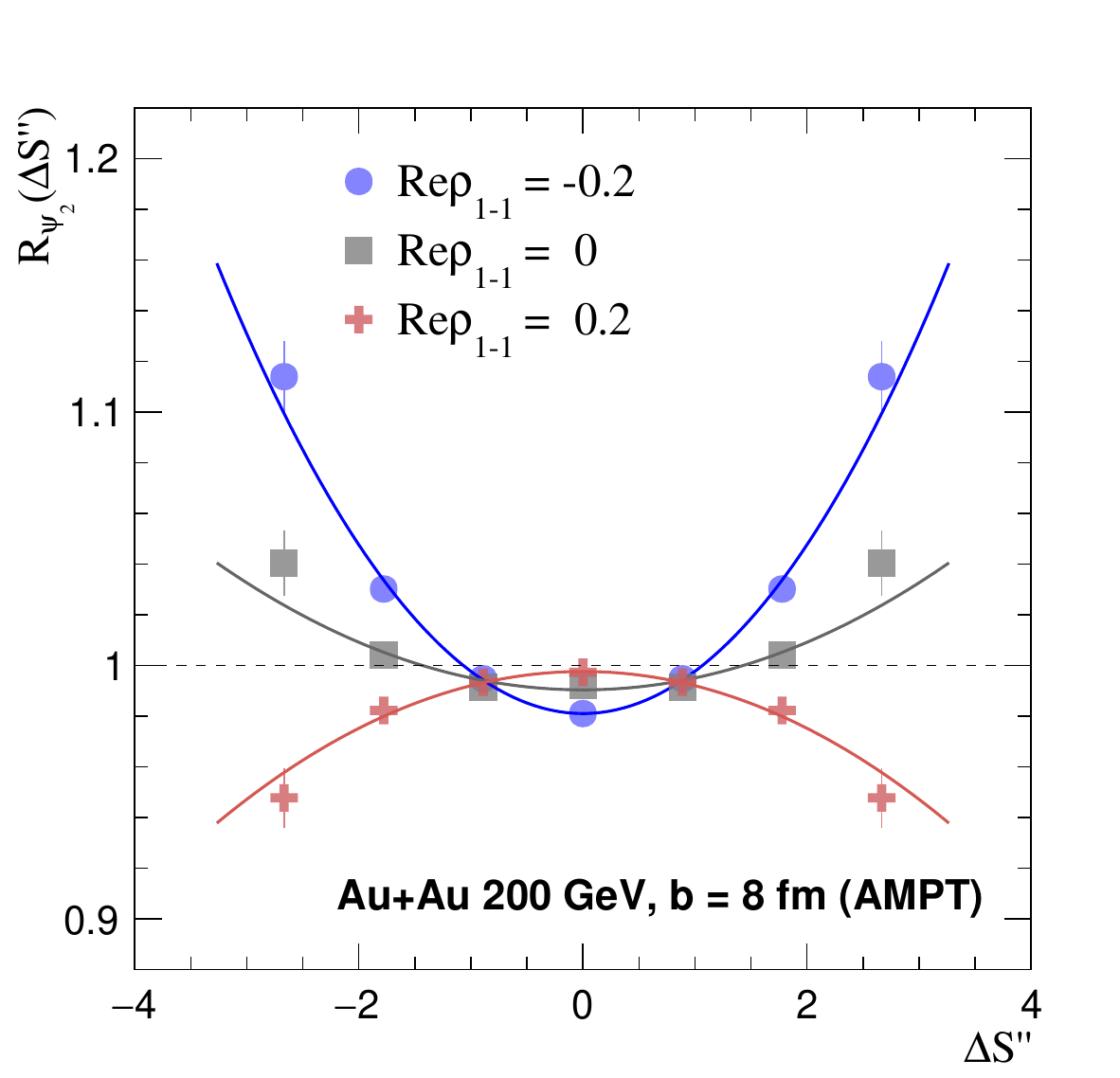}} 
{\includegraphics[width=0.4\linewidth]{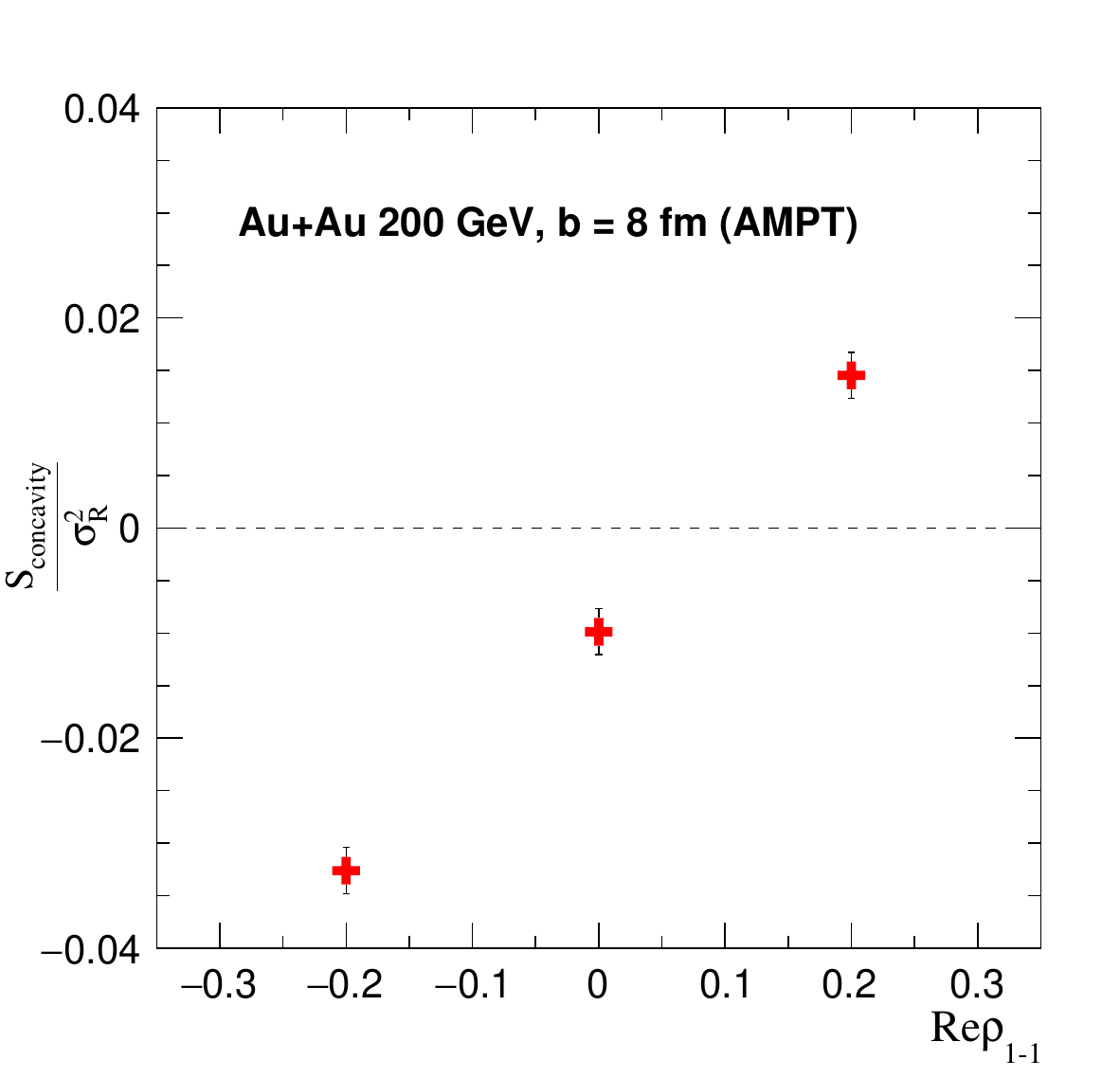}} 
\caption{(Left) AMPT calculations of the $R_{\Psi_2}(\Delta S'')$ distribution with $\mathrm{Re}\,\rho_{1-1}=$ 0.2, 0, and $-0.2$ in Au+Au collisions at $\sqrt{s_{NN}}=200$ GeV with an impact parameter of 8 fm. $\rho_{00}$ is set to $1/3$. The distributions are symmetrized around $\Delta S''=$0. (Right) $S_{\rm concavity}/\sigma_{R}^2$ extracted using Gaussian fits vs $\mathrm{Re}\,\rho_{1-1}$. }
	\label{fig:RCor_AMPT}
\end{figure*}
Figure~\ref{fig:RCor_AMPT} shows the AMPT calculations of the $R_{\Psi_2}(\Delta S'')$ distribution with $\mathrm{Re}\,\rho_{1-1}= -0.2$, 0, and 0.2 in Au+Au collisions at $\sqrt{s_{NN}}=200$ GeV with an impact parameter of 8 fm. $\rho_{00}$ is set to $1/3$.
In contrast to the toy-model results, the $R_{\Psi_2}(\Delta S'')$ distribution manifests a concave shape at zero $\mathrm{Re}\,\rho_{1-1}$, which suggests a non-CME background due to the positive $v_2^\rho$ value and the local charge conservation effect. Figure~\ref{fig:RCor_AMPT} (right) presents the $S_{\rm concavity}/\sigma_{R}^2$ values retrieved using Gaussian fits, showing a linear dependence on Re$\,\rho_{1-1}$, similar to the toy-model outcomes. 
Thus, the AMPT calculations further confirm the background contribution of $\rho$-meson spin coherence to the $R_{\Psi_2}(\Delta S)$ correlator.
Note that a quantitative estimation depends on model details, as evidenced by the difference between the toy model and the AMPT.

\section{Signed balance functions}
Signed balance functions probe the CME by examining the momentum ordering between two charged particles, based on the following quantity~\cite{20Tang:2019pbl,24Choudhury:2021jwd} 
\begin{eqnarray}
\Delta B_y  &\equiv& \frac{N_{y(+-)}-N_{y(++)}}{N_+} - \frac{N_{y(-+)}-N_{y(--)}}{N_-} \notag \\
 & &- \frac{N_{y(-+)}-N_{y(++)}}{N_+} + \frac{N_{y(+-)}-N_{y(--)}}{N_-} \notag \\
&=& \frac{N_+ + N_-}{N_+N_-}[N_{y(+-)} - N_{y(-+)}],
\label{Eq:Delta_By}	
\end{eqnarray}
where $N_{y(\alpha\beta)}$ is the number of pairs in which particle $\alpha$ is ahead of particle $\beta$ along the $y$ axis ($p_y^\alpha > p_y^\beta$) in an event. Similarly, $\Delta B_x$ can be constructed as a reference along the $x$ axis,
\begin{equation}
\Delta B_x = \frac{N_+ + N_-}{N_+N_-}[N_{x(+-)} - N_{x(-+)}].
\label{Eq:Delta_Bx}	
\end{equation}

The CME-induced charge separation along the $y$ axis enhances the width of the $\Delta B_y$ distribution, but not that of the $\Delta B_x$ distribution. Therefore, the final observable is the ratio
\begin{equation}
r \equiv \sigma(\Delta B_y)/\sigma(\Delta B_x).
\end{equation}
$r$ can be calculated in either the laboratory frame ($r_{\rm lab}$) or the pair's rest frame ($r_{\rm rest}$). Since the extra sensitivity in $r_{\rm rest}$ is a higher-order effect requiring substantially more statistics and computing resources, we only focus on $r_{\rm lab}$ in this work. 
As suggested in Refs.~\cite{24Choudhury:2021jwd,DShen:plb}, we use a more straightforward definition based on the difference instead of the ratio to simplify the analytical derivation,
\begin{equation}
\Delta \sigma^2(\Delta B) \equiv \sigma^2(\Delta B_y) - \sigma^2(\Delta B_x), \label{Eq:BF_delta0}
\end{equation}
where (see the appendix of Ref.~\cite{DShen:plb} for details) 
\begin{eqnarray}
\sigma^2(\Delta B_y) &\approx& \frac{64M^2}{\pi^4} \left ( \frac{4}{9M} + 1 \right )\sigma^2(\Delta S_{\rm real}), \label{Eq:SigmaBy}\\
\sigma^2(\Delta B_x) &\approx& \frac{64M^2}{\pi^4} \left ( \frac{4}{9M} + 1 \right )\sigma^2(\Delta S^\perp_{\rm real}). \label{Eq:SigmaBx} 
\end{eqnarray}
Here, $M$ is the total multiplicity of all charged particles.
According to Eqs.~(\ref{Eq:VarYReal}) and (\ref{Eq:VarXReal}), we can relate  $\sigma(\Delta B_y)$ and $\sigma(\Delta B_x)$ to the key components of the $R_{\Psi_2}(\Delta S)$ correlator, and thus rewrite $\Delta \sigma^2(\Delta B)$ as
\begin{equation}
\Delta \sigma^2(\Delta B) \approx c_1+c_2\left[ \: \frac{3}{2}(\rho_{00}-\frac{1}{3}) - \mathrm{Re}\,\rho_{1-1} \: \right], \label{Eq:BF_delta}
\end{equation}
where $c_1$ and $c_2$ are constant coefficients that depend on the $\rho$-meson spectrum, $v_2^\rho$, and $v_2^\pi$.

\begin{figure}[htbp]
	\centering
{\includegraphics[scale=0.33]{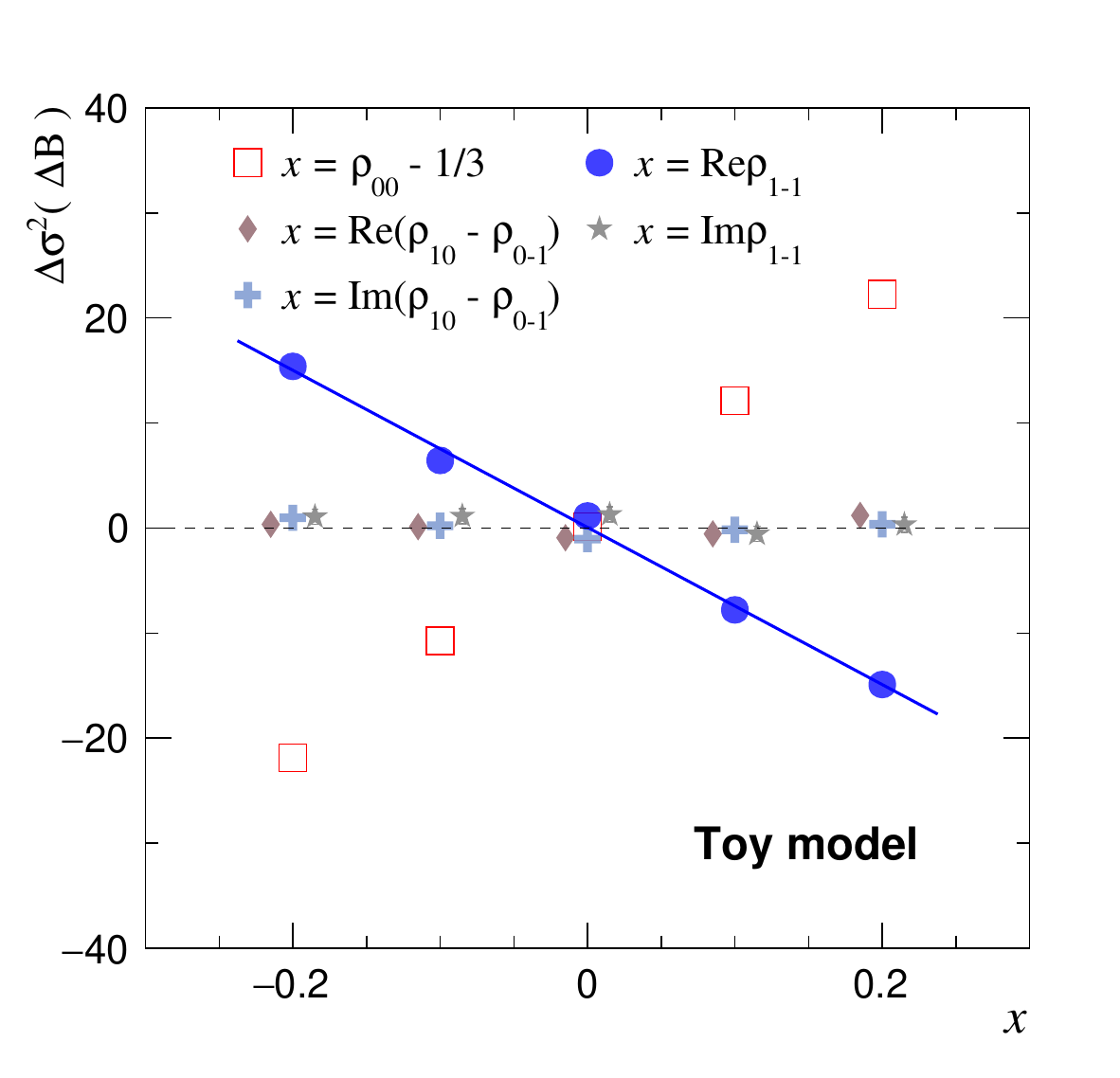}}	
\caption{Toy model simulations of $\Delta \sigma^2(\Delta B)$ vs various elements of the $\rho$-meson spin density matrix: $\rho_{00}-1/3$, ${\rm Re}(\rho_{10}-\rho_{0-1})$, ${\rm Im}(\rho_{10}-\rho_{0-1})$, ${\rm Re}\,\rho_{1-1}$, and ${\rm Im}\,\rho_{1-1}$.
$v_2^\rho$ is set to zero.  The solid line represents the linear fit to the case $x = {\rm Re}\,\rho_{1-1}$.}
	\label{fig:BF_Delta}
\end{figure}
Figure~\ref{fig:BF_Delta} shows the toy model simulations of $\Delta \sigma^2(\Delta B)$ as a function of each element of the $\rho$-meson spin density matrix,
with the same settings as those for the $\Delta\gamma_{112}$ calculations. 
Among the off-diagonal elements, only Re$\,\rho_{1-1}$ has nonzero contributions to $\Delta \sigma^2(\Delta B)$, exerting an effect opposite to that of $\rho_{00}$. 
The function forms of Eqs.~(\ref{Eq:gamma_lab}), (\ref{Eq:RCor_delta}), and (\ref{Eq:BF_delta}) are very similar to each other, as manifested in Figs.~\ref{fig:Gamma_toy}, \ref{fig:deltaR_toy}, and \ref{fig:BF_Delta}. Therefore, this further consolidates that these observables have the same sensitivity to backgrounds.

\begin{figure}[htbp]
	\centering
{\includegraphics[scale=0.33]{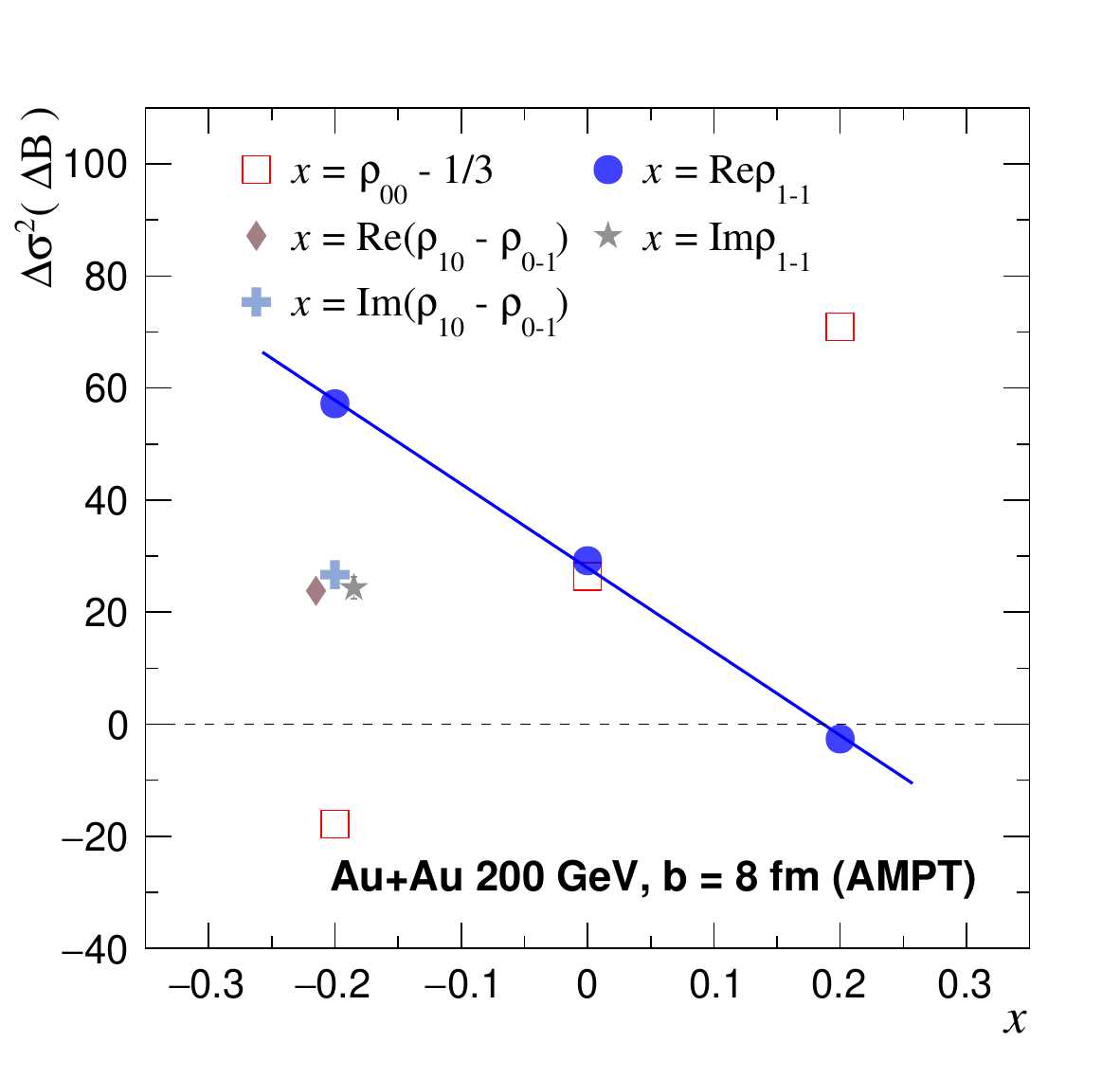}} 	
\caption{AMPT calculations of $ \Delta \sigma^2(\Delta B)$ vs various elements of the $\rho$-meson spin density matrix: $\rho_{00}-1/3$, ${\rm Re}(\rho_{10}-\rho_{0-1})$, ${\rm Im}(\rho_{10}-\rho_{0-1})$, ${\rm Re}\,\rho_{1-1}$, and ${\rm Im}\,\rho_{1-1}$ in Au+Au collisions at $\sqrt{s_{NN}}$ = 200 GeV with an impact parameter of 8 fm. The solid line denotes the linear fit for the case  $x = {\rm Re}\,\rho_{1-1}$.}
	\label{fig:BF_DeltaAMPT}
\end{figure}

Figure~\ref{fig:BF_DeltaAMPT} shows the AMPT calculations of $\Delta \sigma^2(\Delta B)$ in Au+Au collisions at $\sqrt{s_{NN}}$ = 200 GeV with an impact parameter of 8 fm, adopting the same input values of the spin density matrix elements as in Fig.~\ref{fig:Gamma_ampt}.
The linear dependence of $\Delta \sigma^2(\Delta B)$ on $\rho_{00}$ and Re$\,\rho_{1-1}$ is akin to the toy model simulations.
The positive $\Delta \sigma^2(\Delta B)$ at $x=$ 0 may originate from the positive $v_2^\rho$ and the local charge conservation effect.
Quantitatively, the slopes $d \Delta \sigma^2(\Delta B)/d\rho_{00}$ and $d \Delta \sigma^2(\Delta B)/d\mathrm{Re}\,\rho_{1-1}$ rely on the specific $\rho$-meson spectrum.

\section{Summary}
The chiral magnetic effect in high-energy heavy-ion collisions addresses a fundamental question in modern physics: whether parity can be violated in strong interactions. However, the experimental search for the CME has not reached a definitive conclusion because of the complications stemming from the non-CME backgrounds in the observables. In this work, we extend our previous study~\cite{DShen:plb} and demonstrate that not only the global spin alignment ($\rho_{00}$) of vector mesons but also the real part of the off-diagonal element, the Re$\,\rho_{1-1}$, can have a non-negligible contribution to the major CME observables.   

For each of the $\Delta \gamma_{112}$ correlator, the $R_{\Psi_2}(\Delta S)$ correlator, and the signed balance functions, we analytically derive its qualitative dependence on the elements of the vector meson spin density matrix and find that only $\rho_{00}$ and Re$\,\rho_{1-1}$ could yield finite contributions.
The effect of Re$\,\rho_{1-1}$ is comparable in magnitude to that of $\rho_{00}$ but acts in the opposite direction.
We use the $\rho$-meson decays from a toy model and the AMPT model to verify the linear dependence of each CME observable on Re$\,\rho_{1-1}$.
This further confirms the equivalent sensitivity of these three observables to both the CME signal and backgrounds.

The nonzero off-diagonal components in the spin density matrix, known as spin coherence, could have non-CME origins, such as local spin alignment~\cite{30Xia:2020tyd} and short-range spin-spin correlations~\cite{Lv:2024uev}. Hence, the Re$\,\rho_{1-1}$ of vector mesons represents a physics background to CME observables involving the decay products. On the other hand, Eqs.~(\ref{Eq:gamma_lab}), (\ref{Eq:RCor_delta}), and (\ref{Eq:BF_delta}) are not necessarily unidirectional. As noted in Ref.~\cite{Shen:2024xhp}, global spin alignment may also arise from the CME. For instance, the CME induces charge separation of $\pi^+$ and $\pi^-$, some of which later coalesce into $\rho$ mesons with $\rho_{00} \neq 1/3$.
Similarly, we cannot rule out the possibility that a finite Re$\,\rho_{1-1}$ stems from the CME. The interaction between the CME and the elements of the spin density matrix warrants further theoretical investigation.

\section*{Acknowledgement}
Zhiyi Wang and Jinhui Chen are supported, in part, by the National Key Research and Development Program of China under Contract No. 2022YFA1604900, by the National Natural Science Foundation of China (NSFC) under Contract No. 12025501 and the Natural Science Foundation of Shanghai under Contract No. 23JC1400200.
Diyu Shen is supported by the NSFC under Contract No. 12205050.
Gang Wang is supported by the U.S. Department of Energy under Grant No. DE-FG02-88ER40424 and by the NSFC under Contract No.1835002.
Aihong Tang is supported by the US Department of Energy under Grants No. DE-AC02-98CH10886, DE-FG02-89ER40531.

\bibliographystyle{elsarticle-num-names}
\bibliography{mybib}

\end{document}